\documentclass[12pt]{article}
\usepackage{amsmath}
\usepackage[unicode,bookmarks,bookmarksopen,bookmarksopenlevel=2,colorlinks,linkcolor=blue,citecolor=green]{hyperref}

\def\comment#1{}
\input{tcilatex}

\begin{document}

\title{Classification of integrable hydrodynamic chains and generating
functions of conservation laws}
\author{Maxim V. Pavlov}
\date{}
\maketitle

\begin{abstract}
New approach to classification of integrable hydrodynamic chains is
established. Generating functions of conservation laws are classified by the
method of hydrodynamic reductions. $N$ parametric family of explicit
hydrodynamic reductions allows to reconstruct corresponding hydrodynamic
chains. Plenty new hydrodynamic chains are found.
\end{abstract}

\tableofcontents

\textit{keywords}: Riemann surface, hydrodynamic chain, hydrodynamic type
system, generalized hodograph method.


\section{Introduction}

The first integrable hydrodynamic chain%
\begin{equation}
A_{t}^{k}=A_{x}^{k+1}+kA^{k-1}A_{x}^{0}\text{, \ \ \ \ \ }k=0,1,2,...
\label{bm}
\end{equation}%
was derived in \textbf{\cite{Benney}}. The integrability of this
hydrodynamic chain was developed in a set of publications (\textbf{\cite%
{Gibbons}}, \textbf{\cite{Gib+Tsar}}, \textbf{\cite{KM}}, \textbf{\cite{Zakh}%
}). The ``integrability'' means the existence of infinitely many
conservation laws%
\begin{equation}
\partial _{t}\mathbf{H}_{k}(A^{0},A^{1},...,A^{k})=\partial
_{x}G_{k}(A^{0},A^{1},...,A^{k+1})\text{, \ \ \ \ \ }k=0,1,2...  \label{con}
\end{equation}%
and infinitely many commuting flows (see \textbf{\cite{KM}})%
\begin{equation*}
A_{t^{m}}^{k}=[kA^{k+n-1}\partial _{x}+n\partial _{x}A^{k+n-1}]\frac{%
\partial \mathbf{H}_{m+1}}{\partial A^{n}}\text{, \ \ \ \ \ }k,n,m=0,1,2,...,
\end{equation*}%
where $x\equiv t^{0}$, $t\equiv t^{1}$, $\mathbf{H}_{0}=A^{0}$, $\mathbf{H}%
_{1}=A^{1}$, $\mathbf{H}_{2}=A^{2}+(A^{0})^{2}$, $\mathbf{H}%
_{3}=A^{3}+3A^{0}A^{1}$, ...

The first commuting flow is%
\begin{equation}
A_{y}^{k}=A_{x}^{k+2}+A^{0}A_{x}^{k}+(k+1)A^{k}A_{x}^{0}+kA^{k-1}A_{x}^{1}%
\text{, \ \ \ \ \ }k=0,1,2,...,  \label{sh}
\end{equation}%
where $y\equiv t^{2}$. Eliminating the moments $A^{1}$ and $A^{2}$ from the
equations%
\begin{equation*}
A_{t}^{0}=A_{x}^{1}\text{, \ \ \ \ \ }A_{t}^{1}=\partial
_{x}[A^{2}+(A^{0})^{2}/2]\text{, \ \ \ \ \ \ }A_{y}^{0}=\partial
_{x}[A^{2}+(A^{0})^{2}]
\end{equation*}%
one can obtain the Khohlov--Zabolotzkaya equation%
\begin{equation}
A_{tt}^{0}=\partial _{x}[A_{y}^{0}-A^{0}A_{x}^{0}],  \label{kz}
\end{equation}%
which is the dispersionless limit of the KP equation. The compatibility
condition $\partial _{t}(p_{y})=\partial _{y}(p_{t})$ of two generating
functions of conservation laws for the Benney hydrodynamic chain (\textbf{%
\ref{bm}})%
\begin{equation}
p_{t}=\partial _{x}\left( \frac{p^{2}}{2}+A^{0}\right) .  \label{serv}
\end{equation}%
and for its first commuting flow (\textbf{\ref{sh}})%
\begin{equation}
p_{y}=\partial _{x}\left( \frac{p^{3}}{3}+A^{0}p+A^{1}\right)  \label{third}
\end{equation}%
yields (\textbf{\ref{kz}}). Thus, the integrability of this equation is
equivalent the integrability of the Benney hydrodynamic chain (see \textbf{%
\cite{Bogdan}}, \textbf{\cite{Gib+Kod}}, \textbf{\cite{Kod+water}}, \textbf{%
\cite{Krich}}). Substituting the series%
\begin{equation}
p=\lambda -\frac{\mathbf{H}_{0}}{\lambda }-\frac{\mathbf{H}_{1}}{\lambda ^{2}%
}-\frac{\mathbf{H}_{2}}{\lambda ^{3}}-...  \label{ryadok}
\end{equation}%
in both above equations one can obtain (\textbf{\ref{con}}) and the infinite
series of conservation laws for (\textbf{\ref{sh}})%
\begin{equation*}
\partial _{y}\mathbf{H}_{k}(A^{0},A^{1},...,A^{k})=\partial
_{x}Q_{k}(A^{0},A^{1},...,A^{k+2})\text{, \ \ \ \ \ }k=0,1,2...
\end{equation*}

Later plenty integrable hydrodynamic chains%
\begin{equation}
A_{t}^{k}=\underset{n=0}{\overset{k+1}{\sum }}F_{n}^{k}(\mathbf{A})A_{x}^{n}%
\text{, \ \ \ \ \ \ }k=0,1,2,...\text{, \ \ \ \ \ \ }\frac{\partial F_{n}^{k}%
}{\partial A^{m}}=0\text{, \ \ }m>k+1  \label{a}
\end{equation}%
and corresponding 2+1 quasilinear equations (see, for instance, (\textbf{\ref%
{kz}})) was found in \textbf{\cite{Fer+Kar}}, \textbf{\cite{FerKarMax}}, 
\textbf{\cite{Fer+Dav}}, \textbf{\cite{FKT}}, \textbf{\cite{Kuper}}, \textbf{%
\cite{Maks+Egor}}, \textbf{\cite{Zakh+multi}}). Recently some of these
integrable hydrodynamic chains were rediscovered (see \textbf{\cite{Blaszak}}%
) and studied in \textbf{\cite{Blaszak}}, \textbf{\cite{Chang}}, \textbf{%
\cite{Chen+Tu}}, \textbf{\cite{Manas}}, \textbf{\cite{Maks+Kuper}}, \textbf{%
\cite{Maks+eps}}).

At this moment we have several tools allowing to find a complete
classification of the integrable hydrodynamic chains (\textbf{\ref{a}}) and
more complicated hydrodynamic chains%
\begin{equation}
A_{t}^{k}=\underset{n=0}{\overset{N_{k}}{\sum }}F_{n}^{k}(\mathbf{A}%
)A_{x}^{n}\text{, \ \ \ \ \ \ }k=0,1,2,...\text{, \ \ \ \ \ \ }\frac{%
\partial F_{n}^{k}}{\partial A^{m}}=0\text{, \ \ }m>M_{k}\text{,}  \label{e}
\end{equation}%
where $N_{k}$ and $M_{k}$ are some integers.

\textbf{1}. The existence of an extra commuting flow (an \textit{external}
method), successfully used for the Egorov hydrodynamic chains (see \textbf{%
\cite{Maks+Egor}});

\textbf{2}. The existence of an extra conservation law (an \textit{external}
method), successfully (an incomplete classification) used for the
Kupershmidt Poisson bracket and for the (an incomplete classification)
Kupershmidt--Manin Poisson bracket (see \textbf{\cite{Kuper}});

\textbf{3}. Vanishing of the Haantjes tensor (an \textit{internal} method),
successfully (unfinished classification) used for the integrable
hydrodynamic chains (\textbf{\ref{a}}) and successfully (complete
classification) for the Kupershmidt--Manin Poisson bracket (see \textbf{\cite%
{Fer+Dav}});

In this paper we establish the \textit{combined} method based on three key
tools, previously successfully utilized in the theory of 2+1 hydrodynamic
type systems and 2+1 quasilinear equations (see \textbf{\cite{Fer+Kar}}, 
\textbf{\cite{FerKarMax}}, \textbf{\cite{FKT}}, \textbf{\cite{Zakh+multi}}).
These are method of hydrodynamic reductions (see \textbf{\cite{Gib+Tsar}}, 
\textbf{\cite{Fer+Kar}}) and the method of pseudopotentials (see \textbf{%
\cite{Fer+Kar}}, \textbf{\cite{Maks+Egor}}, \textbf{\cite{Zakh+multi}})
combined with the ``concept'' of the Gibbons equation (see \textbf{\cite%
{Maks+algebr}}).

The paper is organized in the following order. In the second section the
concept of the Gibbons equation is introduced. In comparison with (\textbf{%
\cite{Gib+Tsar}}) hydrodynamic reductions are described in a conservative
form. In the third section a new approach of the classification of
integrable hydrodynamic chains via their generating functions of
conservation laws is established. In the fourth section explicit
hydrodynamic reductions associated with local Hamiltonian structures are
discussed. In the fifth section the simple (but nontrivial) generalization
of the Benney hydrodynamic chain is investigated. $N$ parametric family of
the Hamiltonian hydrodynamic reductions is found. The corresponding Riemann
surface is constructed. Thus, $N$ series of conservation laws and commuting
flows can be found (see \textbf{\cite{Maks+algebr}}), then infinitely many
particular solutions can be obtained by the generalized hodograph method
(see \textbf{\cite{Tsar}}). In the sixth section another integrable
hydrodynamic chain is described, whose hydrodynamic reductions coincide with
hydrodynamic reductions of the hydrodynamic chain described in the fifth
section. One particular case of this hydrodynamic chain is connected with
2+1 quasilinear equation determined by the ``integrable'' Lagrangian
considered in \textbf{\cite{FKT}}. In the seventh section three different
approaches allowing to construct commuting flows are presented. The Miura
type transformation is used for links between some well-known hydrodynamic
chains and corresponding 2+1 quasilinear systems via generating functions of
conservation laws. In the eight section we briefly mention the general case
of generating functions of conservation laws connected with arbitrary
integrable hydrodynamic chains. This general case is closely connected with
the Hamiltonian hydrodynamic chains considered in \textbf{\cite{Maks+Hamch}}.

\section{The Gibbons equation}

The Benney hydrodynamic chain (\textbf{\ref{bm}}) is connected with the
formal series%
\begin{equation}
\lambda =p+\frac{A^{0}}{p}+\frac{A^{1}}{p^{2}}+\frac{A^{2}}{p^{3}}+...
\label{ryad}
\end{equation}%
by the Gibbons equation (see details below)%
\begin{equation*}
\lambda _{t}-p\lambda _{x}=\frac{\partial \lambda }{\partial p}\left[
p_{t}-\partial _{x}\left( \frac{p^{2}}{2}+A^{0}\right) \right] .
\end{equation*}%
The \textit{method of hydrodynamic reductions} suggested in \textbf{\cite%
{Gib+Tsar}} (and developed in \textbf{\cite{Fer+Kar}}) means the existence
of infinitely many sub-systems (which are called the ``hydrodynamic
reductions'')%
\begin{equation}
r_{t}^{i}=p^{i}(\mathbf{r})r_{x}^{i}\text{, \ \ \ \ \ }i=1,2,...,N,
\label{ri}
\end{equation}%
where all moments $A^{k}$ are functions of $N$ Riemann invariants $r^{k}$.
This hydrodynamic type system must be consistent with the generating
function of conservation laws (\textbf{\ref{serv}}). Then one can obtain%
\begin{equation}
\partial _{i}p=\frac{\partial _{i}A^{0}}{p^{i}-p}.  \label{com}
\end{equation}%
The compatibility conditions $\partial _{i}(\partial _{k}p)=\partial
_{k}(\partial _{i}p)$ yield the \textit{system in involution}%
\begin{equation}
\partial _{i}p^{k}=\frac{\partial _{i}A^{0}}{p^{i}-p^{k}}\text{, \ \ \ \ \ }%
\partial _{ik}A^{0}=2\frac{\partial _{i}A^{0}\partial _{k}A^{0}}{%
(p^{i}-p^{k})^{2}}\text{, \ \ \ \ \ \ }i\neq k,  \label{gt}
\end{equation}%
which we call the \textit{Gibbons--Tsarev system}.

Any hydrodynamic reduction (\textbf{\ref{ri}}) can be written via first
moments $A^{k}$ ($k=0,1,...,N-1$), where all other moments $A^{n}$ ($%
n=N,N+1,...$) are functions of the first $N$ moments $A^{k}$ (see details in 
\textbf{\cite{Gib+Tsar}}).

Also, \textbf{any} hydrodynamic reduction (\textbf{\ref{ri}}) can be written
in the conservative form (see (\textbf{\ref{serv}}))%
\begin{equation}
a_{t}^{i}=\partial _{x}\left( \frac{(a^{i})^{2}}{2}+A^{0}(\mathbf{a})\right) 
\text{, \ \ \ \ \ }i=1,2,...,N,  \label{sym}
\end{equation}%
where the function $A^{0}$ satisfies the Gibbons--Tsarev system (written via
field variables $a^{k}$)%
\begin{eqnarray}
(a^{i}-a^{k})\partial _{ik}A^{0} &=&\partial _{k}A^{0}\partial _{i}\left(
\sum \partial _{n}A^{0}\right) -\partial _{i}A^{0}\partial _{k}\left( \sum
\partial _{n}A^{0}\right) \text{, \ \ }i\neq k,  \notag \\
&&  \label{egt}
\end{eqnarray}%
\begin{equation*}
(a^{i}-a^{k})\frac{\partial _{ik}A^{0}}{\partial _{i}A^{0}\partial _{k}A^{0}}%
+(a^{k}-a^{j})\frac{\partial _{jk}A^{0}}{\partial _{j}A^{0}\partial _{k}A^{0}%
}+(a^{j}-a^{i})\frac{\partial _{ij}A^{0}}{\partial _{i}A^{0}\partial
_{j}A^{0}}=0\text{, \ \ }i\neq j\neq k,
\end{equation*}%
which is a consequence of the compatibility conditions $\partial
_{i}(\partial _{k}p)=\partial _{k}(\partial _{i}p)$%
\begin{equation}
\frac{p-a^{i}}{\partial _{i}A^{0}}\sum \frac{\partial _{in}A^{0}}{p-a^{n}}-%
\frac{p-a^{k}}{\partial _{k}A^{0}}\sum \frac{\partial _{kn}A^{0}}{p-a^{n}}+%
\frac{(a^{i}-a^{k})\partial _{ik}A^{0}}{\partial _{i}A^{0}\partial _{k}A^{0}}%
\left( \sum \frac{\partial _{n}A^{0}}{p-a^{n}}-1\right) =0,  \label{tri}
\end{equation}%
where $\partial _{i}\equiv \partial /\partial a^{i}$ and (cf. (\textbf{\ref%
{com}}))%
\begin{equation}
\partial _{i}p=\frac{\partial _{i}A^{0}}{p-a^{i}}\left( \sum \frac{\partial
_{n}A^{0}}{p-a^{n}}-1\right) ^{-1}.  \label{sis}
\end{equation}

\textbf{Example}: The simplest hydrodynamic reduction is given by $%
A^{0}=\sum \varepsilon _{k}a^{k}$. This is so-called ``waterbag'' reduction
(see \textbf{\cite{Gib+Yu}}, \textbf{\cite{Kod+water}}). Then (\textbf{\ref%
{sis}})%
\begin{equation*}
\partial _{i}p=\frac{\varepsilon _{i}}{p-a^{i}}\left( \sum \frac{\varepsilon
_{n}}{p-a^{n}}-1\right) ^{-1}
\end{equation*}%
can be integrated. The equation of the Riemann surface%
\begin{equation}
\lambda =p-\sum \varepsilon _{k}\ln (p-a^{k}),  \label{water}
\end{equation}%
where $\lambda $ is an integration factor, can be expanded (at the infinity $%
\lambda \rightarrow \infty $, $p\rightarrow \infty $) in the formal series (%
\textbf{\ref{ryad}}), where $A^{k}=\Sigma \varepsilon
_{i}(a^{i})^{k+1}/(k+1) $, if $\Sigma \varepsilon _{i}=0$. If $\Sigma
\varepsilon _{i}\neq 0$, then at first the parameter $\lambda $ in the
equation of the Riemann surface must be re-scaled%
\begin{equation*}
\lambda -\sum \varepsilon _{k}\ln \lambda =p-\sum \varepsilon _{k}\ln
(p-a^{k}).
\end{equation*}%
Then all other moments $A^{k}$ are some non-homogeneous polynomials of $%
\Sigma \varepsilon _{i}(a^{i})^{k}$.

This is not an unique choice. For instance, any hydrodynamic reductions can
be written in a conservative form in infinitely many ways. Let us mention
just two simplest choices here (all other are discussed in \textbf{\cite%
{Maks+algebr}}):%
\begin{eqnarray}
\mathbf{1}\text{. \ \ \ \ }a_{t}^{i} &=&\partial _{x}\left( \frac{(a^{i})^{2}%
}{2}+A^{0}(\mathbf{a,b})\right) \text{, \ \ \ \ \ \ }b_{t}^{i}=\partial
_{x}(a^{i}b^{i})\text{ \ \ \ \ \ }i=1,2,...,N,  \label{ek} \\
&&  \notag \\
\mathbf{2}\text{. \ \ \ \ }a_{t}^{i} &=&\partial _{x}\left( \frac{(a^{i})^{2}%
}{2}+A^{0}\right) \text{, \ \ \ \ \ \ }A_{t}^{0}=\partial _{x}A^{1}(\mathbf{a%
})\text{ \ \ \ \ \ }i=1,2,...,N,  \notag
\end{eqnarray}%
For the first choice, the simplest hydrodynamic reduction is given by $%
A^{0}=\Sigma b^{k}$; for the second choice $A^{1}=\Sigma b^{k}$.

The Riemann invariants are most suitable coordinates to prove such general
properties like the existence of infinitely many hydrodynamic reductions
parameterized by $N$ arbitrary functions of a single variable (see \textbf{%
\cite{Gib+Tsar}}), but they are very inconvenient for a search of particular
and explicit hydrodynamic reductions. Vice versa, it is very difficult to
derive the Gibbons--Tsarev system in the coordinates $a^{k}$ (see (\textbf{%
\ref{sis}})), but explicit particular hydrodynamic reductions can be found
naturally in many cases (see a lot of examples in \textbf{\cite{Maks+Kuper}}%
).

The \textit{phenomenological} algebro-geometric approach for integrability
of symmetric hydrodynamic type systems%
\begin{equation}
a_{t}^{i}=\partial _{x}\psi (a^{1},a^{2},...,a^{N};p)|_{p=a^{i}}  \label{1}
\end{equation}%
was formulated in \textbf{\cite{Maks+algebr}}.

\textbf{Statement 1}: \textit{If the symmetric hydrodynamic type system} (%
\textbf{\ref{1}}) \textit{is integrable, then this system has the generating
function of conservation laws}%
\begin{equation}
p_{t}=\partial _{x}\psi (a^{1},a^{2},...,a^{N};p).  \label{2}
\end{equation}

\textbf{Statement 2}: \textit{Then some function} $\lambda
(a^{1},a^{2},...,a^{N};p)$ \textit{satisfies the Gibbons equation}%
\begin{equation}
\lambda _{t}-\frac{\partial \psi }{\partial p}\lambda _{x}=\frac{\partial
\lambda }{\partial p}[p_{t}-\partial _{x}\psi (\mathbf{a};p)].  \label{3}
\end{equation}

We call the function $\lambda (a^{1},a^{2},...,a^{N};p)$ the equation of the
Riemann surface. This function is a solution of the set of \textit{linear}
PDE's%
\begin{equation*}
A_{i}^{k}\frac{\partial \lambda }{\partial u^{k}}+\frac{\partial \psi }{%
\partial u^{i}}\frac{\partial \lambda }{\partial p}=0,
\end{equation*}%
where the matrix $A_{i}^{k}$\ is given by%
\begin{equation*}
A_{i}^{k}(\mathbf{u};p)=\left( \frac{\partial \psi }{\partial p}|_{p=u^{i}}-%
\frac{\partial \psi }{\partial p}\right) \delta _{i}^{k}+\frac{\partial \psi 
}{\partial u^{i}}|_{p=u^{k}}.
\end{equation*}

The Gibbons equation (\textbf{\ref{3}}) describes a deformation of the
Riemann surface $\lambda (\mathbf{a};p)$. This equation has three
distinguish features:

\textbf{1}. if one fixes $\lambda =\limfunc{const}$ (free parameter), then
one obtains (\textbf{\ref{2}}),

\textbf{2}. if one fixes $p=\limfunc{const}$ (free parameter), then one
obtains the kinetic equation (a collisionless Vlasov equation) written in
so-called Lax form%
\begin{equation*}
\lambda _{t^{1}}=\{\lambda \text{, }\mathbf{\hat{H}}\}=\frac{\partial
\lambda }{\partial x}\frac{\partial \mathbf{\hat{H}}}{\partial p}-\frac{%
\partial \lambda }{\partial p}\frac{\partial \mathbf{\hat{H}}}{\partial x},
\end{equation*}%
where $\mathbf{\hat{H}}=\psi (\mathbf{a};p)$.

\textbf{3}. if one choose coordinates, which are the Riemann invariants
determined by the condition $\partial \lambda /\partial p=0$, then the
corresponding hydrodynamic type system (\textbf{\ref{1}}) can be written in
the diagonal form (cf. (\textbf{\ref{ri}}))%
\begin{equation}
r_{t^{1}}^{i}=\mu ^{i}(\mathbf{r})r_{x}^{i}\text{, \ \ \ \ \ }i=1,2,...,N,
\label{rim}
\end{equation}%
where the characteristic velocities%
\begin{equation*}
\mu ^{i}=\frac{\partial \psi }{\partial p}|_{p=p^{i}}
\end{equation*}%
can be found from the algebraic system $\det A_{i}^{k}(\mathbf{u};p)=0$;
then the corresponding values $p^{i}$ can be expressed via the Riemann
invariants $r^{k}$. In this algebro-geometric construction the Riemann
invariants are the branch points $r^{i}=\lambda |_{\partial \lambda
/\partial p=0}$ of the Riemann surface (exactly as it is in the Whitham
theory, see \textbf{\cite{Dubr}} and \textbf{\cite{Krich}}).

\textbf{Example}: The dispersionless limit of the vector NLS (see (\textbf{%
\ref{ek}}) and \textbf{\cite{Zakh}})%
\begin{equation}
a_{t}^{i}=\partial _{x}\left( \frac{(a^{i})^{2}}{2}+\sum b^{n}\right) \text{%
, \ \ \ \ \ \ \ }b_{t}^{i}=\partial _{x}(a^{i}b^{i})\text{, \ \ \ \ \ \ \ }%
i=1,2,...,N  \label{za}
\end{equation}%
is the first known hydrodynamic reduction of the Benney hydrodynamic chain (%
\textbf{\ref{bm}}) determined by the moment decomposition $A^{k}=\sum
(a^{i})^{k}b^{i}$. In such case the formal series (\textbf{\ref{ryad}})
reduces to the equation of the Riemann surface%
\begin{equation}
\lambda =p+\sum \frac{b^{k}}{p-a^{k}}.  \label{zak}
\end{equation}%
The Zakharov reduction (\textbf{\ref{za}}) written in the Riemann invariants
(\textbf{\ref{rim}}) (see (\textbf{\ref{ri}}))%
\begin{equation*}
r_{t}^{i}=p^{i}(\mathbf{r})r_{x}^{i}\text{, \ \ \ \ \ }i=1,2,...,2N,
\end{equation*}%
where the characteristic velocities $p^{i}(\mathbf{r})$ can be found from ($%
\partial \lambda /\partial p=0$)%
\begin{equation*}
1=\sum \frac{b^{k}}{(p-a^{k})^{2}}
\end{equation*}%
has infinitely many conservation laws, which can be obtained from (\textbf{%
\ref{zak}}) with the aid of the B\"{u}rmann--Lagrange series (see, for
instance, \textbf{\cite{Lavr}} and some details in \textbf{\cite{Maks+algebr}%
}).

The compatibility conditions $\partial _{i}(\partial _{k}p)=\partial
_{k}(\partial _{i}p)$ must be valid \textit{identically} for any symmetric
hydrodynamic type system (\textbf{\ref{1}}), but a whole set (parameterized
by $N$ arbitrary functions of a single variable) of such symmetric
hydrodynamic type systems (for each fixed function $\psi (\mathbf{a};p)$) is 
\textit{described} by these compatibility conditions $\partial _{i}(\partial
_{k}p)=\partial _{k}(\partial _{i}p)$.

Thus, the problem of the description of semi-Hamiltonian symmetric
hydrodynamic type system is the problem of the classification of integrable
hydrodynamic chains.

\section{The Gibbons--Tsarev system}

The \textbf{main claim} of this paper is that \textit{the classification of
integrable hydrodynamic chains} (\textbf{\ref{e}}) \textit{is equivalent to
the classification of the generating functions of conservation laws} (%
\textbf{\ref{2}}) \textit{by the method of hydrodynamic reductions (see} 
\textbf{\cite{Gib+Tsar}}, \textbf{\cite{Fer+Kar}}).

In this section we derive integrability conditions, which are a nonlinear
PDE's system in involution. This \textit{extended Gibbons--Tsarev system}
generalizes the Gibbons--Tsarev system obtained in \textbf{\cite{Gib+Tsar}}
and describes a \textit{set} of generating functions of conservation laws
together with their hydrodynamic reductions parameterized by $N$ arbitrary
functions of a single variable.

The simplest case is%
\begin{equation}
p_{t}=\partial _{x}\psi (u;p).  \label{b}
\end{equation}

Let us first introduce the new notations%
\begin{equation}
f_{i}\equiv f(u,p^{i},p)=\frac{\psi _{u}}{\psi _{p}|_{p=p^{i}}-\psi _{p}}%
\text{, \ \ }f_{ik}\equiv f(u,p^{i},p^{k})=\frac{\psi _{u}|_{p=p^{k}}}{\psi
_{p}|_{p=p^{i}}-\psi _{p}|_{p=p^{k}}}\text{, \ }i\neq k,  \label{7}
\end{equation}%
\begin{equation}
\varphi _{ik}(u,p^{i},p^{k},p)\equiv \frac{f_{ik}\partial
_{p^{k}}f_{k}-f_{ki}\partial _{p^{i}}f_{i}+f_{i}\partial
_{p}f_{k}-f_{k}\partial _{p}f_{i}-\partial _{u}(f_{i}-f_{k})}{f_{i}-f_{k}}%
\text{, \ \ \ \ \ }i\neq k.  \label{8}
\end{equation}%
Differentiating the generating function of conservation laws (\textbf{\ref{b}%
}) with respect to the Riemann invariants $r^{i}$ (see (\textbf{\ref{rim}}%
)), one obtains%
\begin{equation}
\partial _{i}p=f_{i}\partial _{i}u.  \label{4}
\end{equation}%
If $p=p^{k}$ ($k\neq i$), then (\textbf{\ref{4}}) reduces to%
\begin{equation}
\partial _{i}p^{k}=f_{ik}\partial _{i}u.  \label{5}
\end{equation}%
The compatibility condition $\partial _{k}(\partial _{i}p)=\partial
_{i}(\partial _{k}p)$ yields%
\begin{equation}
\partial _{ik}u=\varphi _{ik}\partial _{i}u\partial _{k}u.  \label{6}
\end{equation}%
Thus, the nonlinear PDE's system (\textbf{\ref{5}}), (\textbf{\ref{6}}) is
in involution iff the functions $\varphi _{ik}$ \textit{do \textbf{not}
depend on} $p$ and the compatibility conditions $\partial _{j}(\partial
_{i}p^{k})=\partial _{i}(\partial _{j}p^{k})$ and $\partial _{j}(\partial
_{ik}u)=\partial _{i}(\partial _{jk}u)$ are fulfilled \textit{identically}.
We call the system (\textbf{\ref{5}}), (\textbf{\ref{6}}) the
Gibbons--Tsarev system (cf. (\textbf{\ref{gt}})).

However, in general case the functions $\varphi _{ik}$ \textit{depend on} $p$%
. The compatibility conditions $\partial _{k}(\partial _{i}p^{j})=\partial
_{i}(\partial _{k}p^{j})$ yields (cf. (\textbf{\ref{6}}))%
\begin{equation}
\partial _{ik}u=\bar{\varphi}_{ik}\partial _{i}u\partial _{k}u,  \label{11}
\end{equation}%
where (cf. (\textbf{\ref{8}}))%
\begin{equation*}
\bar{\varphi}_{ik}(u,\mathbf{p})\equiv \varphi
_{ik}(u,p^{i},p^{k},p)|_{p=p^{j}}\text{, \ \ \ \ \ \ }i\neq j\neq k.
\end{equation*}%
Thus, the Gibbons--Tsarev system is determined by (\textbf{\ref{5}}) and (%
\textbf{\ref{11}}).

Finally, one must check the compatibility conditions $\partial _{j}(\partial
_{ik}u)=\partial _{i}(\partial _{jk}u)$%
\begin{equation*}
\partial _{u}(\varphi _{ik}-\varphi _{jk})+\varphi _{ij}(\varphi
_{ik}-\varphi _{jk})+f_{ji}\partial _{p^{i}}\varphi _{ik}+f_{jk}\partial
_{p^{k}}\varphi _{ik}+f_{j}\partial _{p}\varphi _{ik}=f_{ij}\partial
_{p^{j}}\varphi _{jk}+f_{ik}\partial _{p^{k}}\varphi _{jk}+f_{i}\partial
_{p}\varphi _{jk}
\end{equation*}%
following from (\textbf{\ref{6}}), and the compatibility conditions $%
\partial _{j}(\partial _{ik}u)=\partial _{i}(\partial _{jk}u)$ following
from (\textbf{\ref{11}}). Such over-determined nonlinear PDE system is said
to be the \textit{extended} Gibbons--Tsarev system.

Let me emphasize that the extended Gibbons--Tsarev system is a system on the
sole function $\psi (u;p)$ only. \textit{The general solution of this system
yields a classification of integrable hydrodynamic chains}.

\textbf{Remark}: Any $N$ hydrodynamic reduction (\textbf{\ref{rim}}) can be
written in the conservative form (see (\textbf{\ref{b}}))%
\begin{equation}
a_{t}^{i}=\partial _{x}\psi (u(\mathbf{a});a^{i})\text{, \ \ \ \ \ \ \ \ }%
i=1,2,...,N.  \label{f}
\end{equation}%
If the function $\psi (u(\mathbf{a});p)$ determines the integrable
hydrodynamic type system (\textbf{\ref{f}}), then the compatibility
conditions $\partial _{i}(\partial _{k}p)=\partial _{k}(\partial _{i}p)$
satisfy \textit{identically}, where $\partial _{k}\equiv \partial /\partial
a^{k}$ and (cf. (\textbf{\ref{sis}}))%
\begin{equation}
\partial _{i}p=\frac{\psi _{u}\partial _{i}u}{\psi _{p}|_{p=a^{i}}-\psi _{p}}%
\left[ 1+\sum \frac{\psi _{u}|_{p=a^{k}}\partial _{k}u}{\psi
_{p}|_{p=a^{k}}-\psi _{p}}\right] ^{-1}.  \label{g}
\end{equation}%
If the function $\psi (u;p)$ determines the integrable hydrodynamic chain,
then the compatibility conditions $\partial _{i}(\partial _{k}p)=\partial
_{k}(\partial _{i}p)$ describe its $N$ component hydrodynamic reductions
parameterized by $N$ arbitrary functions of a single variable. If the
function $\psi (u;p)$ is unknown, then the compatibility conditions $%
\partial _{i}(\partial _{k}p)=\partial _{k}(\partial _{i}p)$ and the
compatibility conditions $\partial _{j}(\partial _{ik}u)=\partial
_{i}(\partial _{jk}u)$ yield the extended Gibbons--Tsarev system.

Let us finally emphasize: the Gibbons--Tsarev system describes $N$ component
hydrodynamic reductions parameterized by $N$ arbitrary functions of a single
variable for any a priori given function $\psi (\mathbf{u};p)$; the extended
Gibbons--Tsarev system describes all possible functions $\psi (\mathbf{u};p)$
and their $N$ component hydrodynamic reductions parameterized by $N$
arbitrary functions of a single variable.

\section{Explicit Hamiltonian hydrodynamic reductions}

Suppose the hydrodynamic reductions (\textbf{\ref{sym}}) of the Benney
hydrodynamic chain (\textbf{\ref{bm}}) (simultaneously, they are
hydrodynamic reductions of the Khohlov--Zabolotzkaya equation (\textbf{\ref%
{kz}})) are Hamiltonian (see details in \textbf{\cite{Dubr+Nov}})%
\begin{equation*}
a_{t}^{i}=\frac{1}{\varepsilon _{i}}\partial _{x}\frac{\partial \mathbf{h}}{%
\partial a^{i}},
\end{equation*}%
where $\varepsilon _{i}$ are arbitrary constants. Then the Hamiltonian
density $\mathbf{h}=\Sigma \varepsilon _{k}(a^{k})^{3}/6+f(\Delta )$, where $%
\Delta =\Sigma \varepsilon _{k}a^{k}$ and $f^{\prime }(\Delta )=A^{0}(\Delta
)$. Substitution $A^{0}(\Delta )$ in (\textbf{\ref{tri}}) yields the choice $%
A^{0}(\Delta )=\Delta $ only (up to\ an insufficient constant factor). This
is exactly so-called ``waterbag'' hydrodynamic reduction (see (\textbf{\ref%
{water}})).

Thus, the \textbf{main claim} of this section is that the Hamiltonian
hydrodynamic reductions (cf. (\textbf{\ref{f}}))%
\begin{equation*}
a_{t}^{i}=\partial _{x}\psi (\mathbf{u};a^{i})
\end{equation*}%
for any hydrodynamic chain determined by the the generating function of
conservation laws (see (\textbf{\ref{3}}) and (\textbf{\ref{b}}))%
\begin{equation*}
p_{t}=\partial _{x}\psi (\mathbf{u};p),
\end{equation*}%
one can seek in the form%
\begin{equation}
a_{t}^{i}=\partial _{x}\left( \delta _{i}\frac{\partial \mathbf{h}}{\partial
a^{i}}+\gamma _{i}\underset{k\neq i}{\sum }\gamma _{k}\frac{\partial \mathbf{%
h}}{\partial a^{k}}\right) ,  \label{gam}
\end{equation}%
where $\delta _{i}$ and $\gamma _{i}$ are some constants (see examples in 
\textbf{\cite{Maks+algebr}}).

The second step is a reconstruction of the Riemann surface determined by the
equation $\lambda (\mathbf{u};p)$ by virtue of (\textbf{\ref{gam}}) and a
computation of the moments $A^{k}=\Sigma f_{n}^{k}(a^{n})$, where the
functions $f_{n}^{k}(a^{n})$ from the expansion of the equation of the
Riemann surface at infinity ($\lambda \rightarrow \infty ,p\rightarrow
\infty $). Couple of examples are considered below.

\section{Generalized Benney hydrodynamic chain}

In this section we restrict our consideration on the first simple and very
important particular case (cf. (\textbf{\ref{serv}}))%
\begin{equation}
p_{t}=\partial _{x}[U(p)+u].  \label{c}
\end{equation}

\textbf{Theorem}: $N$ \textit{component hydrodynamic reductions} (\textbf{%
\ref{rim}})%
\begin{equation*}
r_{t}^{i}=U^{\prime }(p^{i})r_{x}^{i}
\end{equation*}%
\textit{are described by solutions of the Gibbons--Tsarev system} (cf. (%
\textbf{\ref{gt}}))%
\begin{eqnarray}
\partial _{ik}u &=&\frac{2\alpha U^{\prime }(p^{i})U^{\prime }(p^{k})+\beta
\lbrack U^{\prime }(p^{i})+U^{\prime }(p^{k})]+2\gamma }{(U^{\prime
}(p^{i})-U^{\prime }(p^{k}))^{2}}\partial _{i}u\partial _{k}u,  \notag \\
&&  \label{i} \\
\partial _{i}p &=&\frac{\partial _{i}u}{U^{\prime }(p^{i})-U^{\prime }(p)}%
\text{, \ \ \ \ \ \ \ \ }\partial _{i}p^{k}=\frac{\partial _{i}u}{U^{\prime
}(p^{i})-U^{\prime }(p^{k})},  \notag
\end{eqnarray}%
\textit{where the function} $U(p)$ \textit{is a solution of the second order
ODE with respect to the independent variable} $p$%
\begin{equation}
U^{\prime \prime }(p)=\alpha U^{\prime ^{2}}(p)+\beta U^{\prime }(p)+\gamma .
\label{ode}
\end{equation}

The compatibility conditions $\partial _{j}(\partial _{ik}A^{0})=\partial
_{i}(\partial _{jk}A^{0})$ are identically satisfied for any constants $%
\alpha $, $\beta $, $\gamma $. In general case ($\alpha \neq 0$) the
function $U(p)$ can be written in the parametric form only%
\begin{equation*}
U=\frac{1}{\alpha (q_{1}-q_{2})}[q_{1}\ln (q-q_{1})-q_{2}\ln (q-q_{2})]\text{%
, \ \ \ \ \ }p=\frac{1}{\alpha (q_{1}-q_{2})}[\ln (q-q_{1})-\ln (q-q_{2})].
\end{equation*}%
If $\alpha =0$, but $\beta \neq 0$, then without lost of generality $U=e^{p}$%
; if $\beta =0$, then $U=p^{2}/2$; if $\alpha \neq 0$, but $\beta =0$ and $%
\gamma =0$, then $U=\ln p$; if $\alpha \neq 0$, but $\beta =0$, then $U=\ln
\sinh p$ (or $U=\ln \cosh p$).

\textbf{The first step in the reconstruction of integrable hydrodynamic
chains is an extracting of \textit{any} explicit hydrodynamic reduction}
(see (\textbf{\ref{f}}))%
\begin{equation}
a_{t}^{i}=\partial _{x}[U(a^{i})+u(\mathbf{a})].  \label{d}
\end{equation}%
Then the compatibility conditions $\partial _{i}(\partial _{k}p)=\partial
_{k}(\partial _{i}p)$ yields (cf. (\textbf{\ref{egt}}), (\textbf{\ref{tri}}%
)) (\textbf{\ref{ode}}), two distinct index relationship%
\begin{equation}
\lbrack U^{\prime }(a^{i})-U^{\prime }(a^{k})](\partial _{ik}u+\alpha
\partial _{i}u\partial _{k}u)=\partial _{k}u\partial _{i}\left( \sum
\partial _{n}u\right) -\partial _{i}u\partial _{k}\left( \sum \partial
_{n}u\right) ,  \label{z}
\end{equation}%
and three distinct index relationship%
\begin{equation*}
\lbrack U^{\prime }(a^{i})-U^{\prime }(a^{k})]\partial _{j}u\partial
_{ik}u+[U^{\prime }(a^{k})-U^{\prime }(a^{j})]\partial _{i}u\partial
_{jk}u+[U^{\prime }(a^{j})-U^{\prime }(a^{i})]\partial _{k}u\partial
_{ij}u=0,
\end{equation*}%
where (see (\textbf{\ref{g}}))%
\begin{equation}
\partial _{i}p=\frac{\partial _{i}u}{U^{\prime }(a^{i})-U^{\prime }(p)}%
\left( \sum \frac{\partial _{k}u}{U^{\prime }(a^{k})-U^{\prime }(p)}%
+1\right) ^{-1}.  \label{der}
\end{equation}

Following the previous section we are looking for the Hamiltonian
hydrodynamic reductions (\textbf{\ref{d}})%
\begin{equation}
a_{t}^{i}=\partial _{x}[U(a^{i})+u(\mathbf{a})]=\frac{1}{\varepsilon _{i}}%
\partial _{x}\frac{\partial \mathbf{h}}{\partial a^{i}}.  \label{h}
\end{equation}

\textbf{Theorem}: \textit{The Hamiltonian hydrodynamic reductions} (\textbf{%
\ref{h}}) \textit{are determined by the Hamiltonian density} $\mathbf{h}%
=\Sigma \varepsilon _{k}\int U(a^{k})da^{k}+f(\Delta )$, \textit{where} $%
\Delta =\Sigma \varepsilon _{k}a^{k}$, $f^{\prime }(\Delta )=u(\Delta )$ 
\textit{and} $u(\Delta )=\Delta $, \textit{if} $\alpha =0$; $u(\Delta )=\ln
(\Delta )$, \textit{if} $\alpha \neq 0$.

\textbf{Proof}: can be obtained by the substitution (\textbf{\ref{h}}) in (%
\textbf{\ref{z}}).

The equation of the Riemann surface (see (\textbf{\ref{der}})) can be found
in quadratures%
\begin{equation}
d\lambda =\frac{e^{\beta p+2\alpha U(p)}}{u^{\prime }(\Delta )}dp+e^{\beta
p+2\alpha U(p)}\sum \frac{\varepsilon _{k}d(p-a^{k})}{U^{\prime
}(a^{k})-U^{\prime }(p)}  \label{rin}
\end{equation}%
for this Hamiltonian hydrodynamic reduction.

Let us introduce the moments $A^{k}=\Sigma \varepsilon _{n}\int U^{\prime
^{k}}(a^{n})da^{n}$, then the hydrodynamic type system (\textbf{\ref{h}})
can be rewritten as the Hamiltonian hydrodynamic chain%
\begin{equation*}
A_{t}^{k}=\underset{n=0}{\overset{k+2}{\sum }}F_{n}^{k}(\mathbf{A})A_{x}^{n}%
\text{, \ \ \ \ \ \ }k=0,1,2,...
\end{equation*}%
determined by the Hamiltonian density $\mathbf{h}%
(A^{0},A^{1})=A^{1}+f(A^{0}) $ (where $f(A^{0})=(A^{0})^{2}/2$, if $\alpha
=0 $ and $f(A^{0})=A^{0}\ln A^{0}$) and by the \textit{Dorfman} Poisson
bracket (see \textbf{\cite{Dorfman}})%
\begin{equation*}
\{A^{0},A^{0}\}=\sum \varepsilon _{n}\delta ^{\prime }(x-x^{\prime }),
\end{equation*}%
\begin{equation*}
\{A^{k},A^{n}\}=[k(\alpha A^{k+n+1}+\beta A^{k+n}+\gamma A^{k+n-1})\partial
_{x}+n\partial _{x}(\alpha A^{k+n+1}+\beta A^{k+n}+\gamma A^{k+n-1})]\delta
(x-x^{\prime }).
\end{equation*}

For instance, this hydrodynamic chain is%
\begin{equation*}
A_{t}^{0}=\partial _{x}[\alpha A^{2}+\beta A^{1}+\left( \gamma +\sum
\varepsilon _{n}\right) A^{0}]
\end{equation*}%
\begin{equation*}
A_{t}^{k}=(\alpha A^{k+2}+\beta A^{k+1}+\gamma A^{k})_{x}+k(\alpha
A^{k+1}+\beta A^{k}+\gamma A^{k-1})A_{x}^{0}\text{, \ \ \ \ }k=1,2,...,
\end{equation*}%
if the Hamiltonian density is $\mathbf{h}(A^{0},A^{1})=A^{1}+(A^{0})^{2}/2$.

\section{The Gibbons--Tsarev system and hydrodynamic chains}

In this section we consider the second simple and very important particular
case%
\begin{equation}
p_{t}=\partial _{x}[V(p)\upsilon ].  \label{new}
\end{equation}%
Then $N$ component hydrodynamic reductions (\textbf{\ref{rim}})%
\begin{equation*}
r_{t}^{i}=\upsilon V^{\prime }(p^{i})r_{x}^{i}
\end{equation*}%
are compatible with the above generating function of conservation laws iff
(see (\textbf{\ref{4}}))%
\begin{equation*}
\partial _{i}p=V(p)\frac{\partial _{i}\ln \upsilon }{V^{\prime
}(p^{i})-V^{\prime }(p)}.
\end{equation*}%
It is easy to see that under re-scaling%
\begin{equation*}
\frac{dp}{V(p)}=d\tilde{p}\text{, \ \ \ \ \ \ }V^{\prime }(p)=U^{\prime }(%
\tilde{p}),
\end{equation*}%
one get the main formula from the previous example (see (\textbf{\ref{i}})),
where $u=\ln \upsilon $. The corresponding equation is (see (\textbf{\ref%
{ode}}))%
\begin{equation}
VV^{\prime \prime }=\alpha V^{\prime ^{2}}+\beta V^{\prime }+\gamma ,
\label{x}
\end{equation}%
where%
\begin{equation}
V(p)=\exp U(\tilde{p})\text{, \ \ \ \ \ \ \ }dp=\exp U(\tilde{p})d\tilde{p}.
\label{u}
\end{equation}%
In general case ($\alpha \neq 0$) the function $V(p)$ can be written in the
parametric form only%
\begin{equation*}
V(p)=(q-q_{1})^{\frac{q_{1}}{\alpha (q_{1}-q_{2})}}(q-q_{2})^{-\frac{q_{2}}{%
\alpha (q_{1}-q_{2})}}\text{, \ \ \ \ \ \ }p=\frac{1}{\alpha }\int
(q-q_{1})^{\frac{q_{1}}{\alpha (q_{1}-q_{2})}-1}(q-q_{2})^{-\frac{q_{2}}{%
\alpha (q_{1}-q_{2})}-1}dq.
\end{equation*}%
The corresponding degenerate cases (see the above example) are $V(p)=p^{k}$,
where $k$ is an arbitrary constant, $V(p)=e^{p}$,\ $V(p)=\cosh p$ (or $%
V(p)=\sinh p$) and two others given in the implicit form%
\begin{equation*}
p=\int \frac{dV}{\ln V}\text{, \ \ \ \ \ \ \ }p=\int \frac{dV}{\sqrt{\ln V}}%
\text{.}
\end{equation*}

Since the equations describing hydrodynamic reductions of these both
examples (\textbf{\ref{c}}) and (\textbf{\ref{new}}) are equivalent, then
one can recalculate any hydrodynamic reduction of the one example to a
corresponding hydrodynamic reduction of the second example (the Hamiltonian
reductions of the generating function of conservation laws (\textbf{\ref{new}%
}) are found in \textbf{\cite{Maks+algebr}}). The relationship (\textbf{\ref%
{u}}) is nontrivial. For instance, $U(\tilde{p})=\tilde{p}^{2}/2$ for the
Benney hydrodynamic chain (see (\textbf{\ref{serv}})), but $V(p)$ cannot be
expressed via known elementary or special functions. Nevertheless,
hydrodynamic reductions of the Benney hydrodynamic chain can be
recalculated. One can compare the formulas (\textbf{\ref{g}}) for (\textbf{%
\ref{c}}) and (\textbf{\ref{new}})%
\begin{eqnarray*}
\frac{\partial p}{\partial a^{i}} &=&\frac{V(p)\partial \upsilon /\partial
a^{i}}{V^{\prime }(a^{i})-V^{\prime }(p)}\left[ \upsilon +\sum \frac{%
V(a^{k})\partial \upsilon /\partial a^{k}}{V^{\prime }(a^{k})-V^{\prime }(p)}%
\right] ^{-1}, \\
&& \\
\frac{\partial \tilde{p}}{\partial c^{i}} &=&\frac{\partial u/\partial c^{i}%
}{U^{\prime }(c^{i})-U^{\prime }(p)}\left[ 1+\sum \frac{\partial u/\partial
c^{k}}{U^{\prime }(c^{k})-U^{\prime }(p)}\right] ^{-1}.
\end{eqnarray*}%
They are equivalent under the transformation (\textbf{\ref{u}}) and under
the same transformation for the field variables $a^{k}\leftrightarrow c^{k}$%
\begin{equation*}
V(a^{i})=\exp U(c^{i})\text{, \ \ \ \ \ \ \ }da^{i}=\exp U(c^{i})dc^{i}.
\end{equation*}%
The equation of the Riemann surface (\textbf{\ref{rin}}) also can be
recalculated%
\begin{equation*}
d\lambda =(V(p))^{2\alpha -1}\exp [\beta \int \frac{dp}{V(p)}]\left( \frac{%
\upsilon (\Delta )dp}{\upsilon ^{\prime }(\Delta )}+\sum \frac{\varepsilon
_{k}}{V(a^{k})}\frac{V(a^{k})dp-V(p)da^{k}}{V^{\prime }(a^{k})-V^{\prime }(p)%
}\right) ,
\end{equation*}%
where $\Delta =\Sigma \varepsilon _{k}\int da^{k}/V(a^{k})$, $\upsilon
(\Delta )=\exp \Delta $, if $\alpha =0$; $\upsilon (\Delta )=\Delta $, if $%
\alpha \neq 0$.

\textbf{Example}: \textit{The Lagrangian quasilinear equation associated
with an elliptic curve}.

The Lagrangian%
\begin{equation*}
L=\int z_{x}z_{y}z_{t}dxdydt
\end{equation*}%
creates the Euler--Lagrange equation%
\begin{equation}
z_{t}z_{xy}+z_{y}z_{xt}+z_{x}z_{yt}=0,  \label{fkt}
\end{equation}%
which is an integrable 2+1 quasilinear equation (see \textbf{\cite{FKT}}).
The pseudopotentials (cf. (\textbf{\ref{kz}}), (\textbf{\ref{serv}}) and (%
\textbf{\ref{third}}); see also \textbf{\cite{Fer+Kar}}, \textbf{\cite%
{Zakh+multi}}) are%
\begin{equation}
\frac{S_{x}}{z_{x}}=\zeta (\sigma )\text{, \ \ \ \ \ \ }\frac{S_{y}}{z_{y}}%
=\zeta (\sigma )+\frac{\wp ^{\prime }(\sigma )+\varepsilon }{2\wp (\sigma )}%
\text{, \ \ \ \ \ \ }\frac{S_{t}}{z_{t}}=\zeta (\sigma )+\frac{\wp ^{\prime
}(\sigma )-\varepsilon }{2\wp (\sigma )}\text{,}  \label{ps}
\end{equation}%
where $\zeta (\sigma )$ and $\wp ^{\prime }(\sigma )$ are Weiershtrass
elliptic functions ($\zeta ^{\prime }(\sigma )=-\wp (\sigma )$, $\wp
^{\prime ^{2}}(\sigma )=4\wp ^{3}(\sigma )+\varepsilon ^{2}$). Introducing
the new functions $a=z_{x}$, $b=z_{y}$, $c=z_{t}$, (\textbf{\ref{fkt}}) can
be written in the form%
\begin{equation*}
a_{t}=c_{x}\text{, \ \ \ }a_{y}=b_{x}\text{, \ \ \ \ }b_{t}=c_{y}\text{, \ \
\ \ }ca_{y}+ba_{t}+ab_{t}=0.
\end{equation*}%
Correspondingly, (\textbf{\ref{ps}}) can be written as the couple of
generating functions of conservation laws%
\begin{equation}
p_{y}=\partial _{x}\left[ \left( \frac{p}{a}+\frac{\wp ^{\prime }(\sigma
)+\varepsilon }{2\wp (\sigma )}\right) b\right] \text{, \ \ \ \ \ \ }%
p_{t}=\partial _{x}\left[ \left( \frac{p}{a}+\frac{\wp ^{\prime }(\sigma
)-\varepsilon }{2\wp (\sigma )}\right) c\right] ,  \label{y}
\end{equation}%
where $p=S_{x}$ and $\sigma $ can be found from $\zeta (\sigma )=p/a$.

The reciprocal transformation (see \textbf{\cite{Rogers}}, \textbf{\cite%
{Yanenko}})%
\begin{equation*}
dz=adx+bdy+cdt\text{, \ \ \ \ }d\tilde{t}=dt\text{, \ \ \ \ \ }d\tilde{y}=dy
\end{equation*}%
reduces the above Lagrangian to%
\begin{equation*}
\tilde{L}=\int \frac{x_{\tilde{y}}x_{\tilde{t}}}{x_{z}^{2}}dzd\tilde{y}d%
\tilde{t}
\end{equation*}%
and (\textbf{\ref{y}}) to the couple of generating functions of conservation
laws%
\begin{equation}
\tilde{p}_{\tilde{y}}=\partial _{z}\left[ \frac{\wp ^{\prime }(\sigma
)+\varepsilon }{2\wp (\sigma )}b\right] \text{, \ \ \ \ \ \ }\tilde{p}_{%
\tilde{t}}=\partial _{z}\left[ \frac{\wp ^{\prime }(\sigma )-\varepsilon }{%
2\wp (\sigma )}c\right] ,  \label{dva}
\end{equation}%
where $\tilde{p}=\zeta (\sigma )$. However, these generating functions of
conservation laws are considered in this section (see (\textbf{\ref{new}})).
Indeed, the equation (\textbf{\ref{x}})%
\begin{equation*}
V\frac{\partial ^{2}V}{\partial \tilde{p}^{2}}=3\left( \frac{\partial V}{%
\partial \tilde{p}}\right) ^{2}+9\frac{\partial V}{\partial \tilde{p}}+6
\end{equation*}%
has the general solution given in the parametric form%
\begin{equation*}
V(\tilde{p})=\frac{\wp ^{\prime }(\sigma )\pm \varepsilon }{2\wp (\sigma )}%
\text{, \ \ \ \ \ \ }\tilde{p}=\zeta (\sigma ).
\end{equation*}

\section{Commuting flows and 2+1 quasilinear systems}

In this section three different approaches in construction of commuting
flows for any given hydrodynamic chain are presented.

\textbf{1}. Suppose all generating functions of conservation laws (\textbf{%
\ref{b}}) are enumerated. Then generating functions of conservation laws for
commuting flows one should seek in the forms%
\begin{equation}
p_{t^{1}}=\partial _{x}\psi _{1}(u^{1};p)\text{, \ \ \ \ \ }%
p_{t^{2}}=\partial _{x}\psi _{2}(u^{1},u^{2};p)\text{, ...}  \label{t}
\end{equation}%
The compatibility conditions $\partial _{t}(p_{t^{1}})=\partial
_{t^{1}}(p_{t})$, $\partial _{t}(p_{t^{2}})=\partial _{t^{2}}(p_{t})$, ...
allow to reconstruct functions $\psi _{k}$ in quadratures (the corresponding
example is given by the Benney hydrodynamic chain (\textbf{\ref{bm}}) and
the Khohlov--Zabolotzkaya equation (\textbf{\ref{kz}}); see (\textbf{\ref%
{serv}}) and (\textbf{\ref{third}})).

\textbf{Example}: Let us consider the couple of commuting generating
functions of conservation laws (cf. (\textbf{\ref{dva}}))%
\begin{equation*}
p_{y}=\partial _{x}[V(p)b]\text{, \ \ \ \ \ \ }p_{t}=\partial _{x}[W(p)c].
\end{equation*}%
The compatibility condition $\partial _{t}(p_{y})=\partial _{y}(p_{t})$
yields (\textbf{\ref{x}})%
\begin{eqnarray*}
VV^{\prime \prime } &=&(1+\frac{\beta }{\alpha })V^{\prime ^{2}}+[\gamma
-\delta -2\frac{\beta \delta }{\alpha }]V^{\prime }+\frac{\delta }{\alpha }%
(\beta \delta -\alpha \gamma ), \\
&& \\
WW^{\prime \prime } &=&(1+\frac{\delta }{\gamma })W^{\prime ^{2}}+[\alpha
-\beta -2\frac{\beta \delta }{\gamma }]W^{\prime }+\frac{\beta }{\gamma }%
(\beta \delta -\alpha \gamma ),
\end{eqnarray*}%
where 2+1 quasilinear system is ($\alpha ,\beta ,\gamma ,\delta $ are
arbitrary constants)%
\begin{equation*}
b_{t}=\alpha bc_{x}+\beta cb_{x}\text{, \ \ \ \ \ \ \ }c_{y}=\gamma
cb_{x}+\delta bc_{x}.
\end{equation*}

\textbf{2}. Let us replace $u^{1}\rightarrow p(\zeta )$ and $%
t^{1}\rightarrow \tau (\zeta )$ in the first equation (\textbf{\ref{t}})%
\begin{equation}
\partial _{\tau (\zeta )}p(\lambda )=\partial _{x}\psi _{1}(p(\zeta
),p(\lambda )).  \label{oba}
\end{equation}

\textbf{Definition}: \textit{The equation} (\textbf{\ref{oba}}) \textit{is
called the generating function of conservation laws and commuting flows}.

All generating functions of conservation laws (\textbf{\ref{t}}) can be
obtained by the expansion of (\textbf{\ref{oba}}) in the series according
the B\"{u}rmann--Lagrange series of $p(\zeta )$, where $\partial _{\tau
(\zeta )}$ is the formal series, whose coefficients $\partial _{t^{k}}$
enumerate different commuting flows.

\textbf{Example}: The Benney hydrodynamic chain has the generating function
of conservation laws (\textbf{\ref{serv}}) and the generating function of
conservation laws and commuting flows is%
\begin{equation}
\partial _{\tau (\zeta )}p(\lambda )=\partial _{x}\ln [p(\lambda )-p(\zeta
)],  \label{zvezda}
\end{equation}%
where $A_{\tau (\zeta )}^{0}=\partial _{x}p(\zeta )$. Substituting the
expansion (\textbf{\ref{ryadok}}) for $p(\lambda )$ in the above formula,
one can obtain the generating function of commuting flows in the form%
\begin{equation*}
A_{\tau (\zeta )}^{0}=\partial _{x}p(\zeta )\text{, \ \ \ \ }A_{\tau (\zeta
)}^{1}=\partial _{x}\left( \frac{(A^{0})^{2}}{2}+p(\zeta )\right) ,...
\end{equation*}%
Substituting the expansion (\textbf{\ref{ryadok}}) for $p(\lambda )$ in the
above formula, one can obtain the generating functions of conservation laws
in the form (see (\textbf{\ref{serv}}) and (\textbf{\ref{third}}))%
\begin{equation*}
p_{t^{1}}=\partial _{x}\left( \frac{p^{2}}{2}+A^{0}\right) \text{, \ \ \ \ \
\ }p_{t^{2}}=\partial _{x}\left( \frac{p^{3}}{3}+A^{0}p+A^{1}\right) ,...
\end{equation*}

\textbf{3}. Suppose for any given function $\psi (u;p)$ (see (\textbf{\ref{b}%
})) we already know the corresponding Hamiltonian hydrodynamic reductions%
\begin{equation*}
a_{t^{1}}^{i}=\partial _{x}\left( \bar{g}^{ik}\frac{\partial \mathbf{h}_{1}}{%
\partial a^{k}}\right) ,
\end{equation*}%
where $\bar{g}^{ik}$ is a constant non-degenerate symmetric matrix. Then one
can seek the \textit{higher} conservation law density $\mathbf{h}_{2}$ for
this hydrodynamic type system. This conservation law density determines the
higher commuting flow%
\begin{equation*}
a_{t^{2}}^{i}=\partial _{x}\left( \bar{g}^{ik}\frac{\partial \mathbf{h}_{2}}{%
\partial a^{k}}\right) .
\end{equation*}%
The generating function of conservation laws for this hydrodynamic type
system (as well as for the corresponding hydrodynamic chain) can be found by
the replacement $a^{i}\rightarrow p$ (see details in \textbf{\cite%
{Maks+algebr}}). The generating functions of commuting flows is given by%
\begin{equation*}
a_{\tau (\zeta )}^{i}=\partial _{x}\left( \bar{g}^{ik}\frac{\partial p(\zeta
)}{\partial a^{k}}\right) .
\end{equation*}

Let us replace $\partial _{\tau (\zeta )}\rightarrow \partial _{t^{-1}}$
and, correspondingly, $p(\zeta )\rightarrow A^{-1}$, then the generating
function of conservation laws (\textbf{\ref{zvezda}}) can be written in the
form%
\begin{equation}
\tilde{p}_{x}=\partial _{t^{-1}}(e^{\tilde{p}}+A^{-1}),  \label{toda}
\end{equation}%
where the \textit{generating function of the Miura type transformations}
(see \textbf{\cite{Maks+Puas}}) is given by $\tilde{p}=\ln (p-A^{-1})$. This
generating function (\textbf{\ref{toda}}) determines the continuum limit of
the discrete KP hierarchy (see \textbf{\cite{Kod+water}})%
\begin{equation*}
B_{x}^{k}=B_{z}^{k+1}+kB^{k}B_{z}^{0}\text{, \ \ \ \ \ }k=0,1,2,...,
\end{equation*}%
where $z\equiv t^{-1}$ and $B^{0}\equiv A^{-1}$.

\textbf{Remark}: All moments $A^{k}$ are connected with the moments $B^{k}$
by the \textit{Miura type transformations}%
\begin{equation*}
A^{0}=A^{0}(B^{0},B^{1})\text{, \ \ \ }A^{1}=A^{1}(B^{0},B^{1},B^{2})\text{,
\ \ \ \ }A^{2}=A^{2}(B^{0},B^{1},B^{2},B^{3}),...,
\end{equation*}%
which can be obtained by a substitution the generating function of the Miura
type transformations $p=A^{-1}+\exp \tilde{p}$ in (\textbf{\ref{ryad}}) and
a comparison with (see \textbf{\cite{Kod+water}})%
\begin{equation*}
\lambda =e^{\tilde{p}}+B^{0}+B^{1}e^{-\tilde{p}}+B^{2}e^{-2\tilde{p}}+...
\end{equation*}

Let us substitute the expansion $\partial _{\tau (\zeta )}=\partial
_{t^{-1}}+\zeta \partial _{t^{-2}}+...$ and $p(\zeta )\rightarrow
A^{-1}+\zeta A^{-2}+...$ in (\textbf{\ref{zvezda}}). Then the compatibility
condition $\partial _{t^{-2}}(\partial _{t^{1}}p)=\partial _{t^{1}}(\partial
_{t^{-2}}p)$ yields the 2+1 shallow water system (see \textbf{\cite%
{Zakh+multi}})%
\begin{equation}
\partial _{t^{-2}}A^{0}=\partial _{x}A^{-2}\text{, \ \ \ \ \ \ }\partial
_{t^{1}}A^{-1}=\partial _{x}\left( \frac{(A^{-1})^{2}}{2}+A^{0}\right) \text{%
, \ \ \ \ \ \ }\partial _{t^{1}}A^{-2}=\partial _{x}(A^{-1}A^{-2}),
\label{j}
\end{equation}%
where%
\begin{equation*}
\partial _{t^{-2}}p=\partial _{x}\frac{A^{-2}}{A^{-1}-p}.
\end{equation*}%
The compatibility condition $\partial _{t^{-2}}(\partial
_{t^{-1}}p)=\partial _{t^{-1}}(\partial _{t^{-2}}p)$ yields the famous
Boyer--Finley equation%
\begin{equation}
\partial _{t^{-1}}A^{-1}=\partial _{x}\ln A^{-2}\text{, \ \ \ \ \ \ \ }%
\partial _{t^{-1}}A^{-2}=\partial _{t^{-2}}A^{-1}\text{.}  \label{bf}
\end{equation}

In the both case (\textbf{\ref{j}}) and (\textbf{\ref{bf}}) $N$ component
hydrodynamic reductions (\textbf{\ref{ri}}) and (see (\textbf{\ref{rim}}))%
\begin{equation*}
r_{t^{-1}}^{i}=\frac{1}{p^{i}-A^{-1}}r_{x}^{i}\text{, \ \ \ \ \ \ \ \ }%
r_{t^{-2}}^{i}=\frac{A^{-2}}{(p^{i}-A^{-1})^{2}}r_{x}^{i}
\end{equation*}%
are compatible (i.e. $\partial _{t^{1}}(r_{t^{-2}}^{i})=\partial
_{t^{-2}}(r_{t^{1}}^{i})$; see \textbf{\cite{Tsar}}, \textbf{\cite{Fer+Kar}})%
\begin{equation*}
\frac{\partial _{k}p^{i}}{p^{k}-p^{i}}=\frac{\partial _{k}\left( \frac{1}{%
p^{i}-A^{-1}}\right) }{\frac{1}{p^{k}-A^{-1}}-\frac{1}{p^{i}-A^{-1}}}=\frac{%
\partial _{k}\left( \frac{A^{-2}}{(A^{-1}-p^{i})^{2}}\right) }{\frac{A^{-2}}{%
(A^{-1}-p^{k})^{2}}-\frac{A^{-2}}{(A^{-1}-p^{i})^{2}}},
\end{equation*}%
iff the functions $p^{i}$ and $A^{0}$ satisfy to the Gibbons--Tsarev system (%
\textbf{\ref{gt}}), where%
\begin{equation*}
\partial _{i}A^{-1}=\frac{\partial _{i}A^{0}}{p^{i}-A^{-1}}\text{, \ \ \ \ \
\ \ }\partial _{i}\ln A^{-2}=\frac{\partial _{i}A^{0}}{(p^{i}-A^{-1})^{2}}.
\end{equation*}%
It is not easy to verify, if did not take into account, that all these 2+1
quasilinear systems (\textbf{\ref{kz}}), (\textbf{\ref{j}}) and (\textbf{\ref%
{bf}}) are members of the unique Benney hydrodynamic \textit{lattice} (see 
\textbf{\cite{Maks+Kuper}}).

\section{General case}

The \textbf{main statement} of this paper is that all integrable
hydrodynamic chains described by the generating functions of conservation
laws (\textbf{\ref{2}}) can be split into sub-classes according to the
number $M$ of the functions $u^{m}$. The first such a case (\textbf{\ref{b}}%
) was considered in the previous section ($M=1$). The second case ($M=2$) is%
\begin{equation*}
p_{t}=\partial _{x}\psi (u,\upsilon ;p).
\end{equation*}%
Then $N$ component hydrodynamic reductions (\textbf{\ref{rim}}) are
compatible with this generating function if the compatibility conditions $%
\partial _{i}(\partial _{k}p)=\partial _{k}(\partial _{i}p)$ are fulfilled,
where%
\begin{equation*}
\partial _{i}p=\frac{\psi _{u}\partial _{i}u+\psi _{\upsilon }\partial
_{i}\upsilon }{\psi _{p}|_{p=p^{i}}-\psi _{p}}.
\end{equation*}%
To complete this computation, we need some relationship between $u$ and $%
\upsilon $ (except trivial link $\upsilon (u)$). Without lost of generality
we can choose these field variables $u$ and $\upsilon $ as a conservation
law density and a flux, respectively:%
\begin{equation*}
u_{t}=\upsilon _{x}\text{.}
\end{equation*}%
Since the hydrodynamic type system (\textbf{\ref{rim}}) has this
conservation law, then%
\begin{equation*}
\partial _{i}\upsilon =\psi _{p}|_{p=p^{i}}\partial _{i}u,
\end{equation*}%
and (cf. (\textbf{\ref{7}}), (\textbf{\ref{4}}))%
\begin{equation*}
\partial _{i}p=\frac{\psi _{u}+\psi _{\upsilon }\psi _{p}|_{p=p^{i}}}{\psi
_{p}|_{p=p^{i}}-\psi _{p}}\partial _{i}u.
\end{equation*}%
However, in such a case the Gibbons--Tsarev system can be derives from the
extended compatibility conditions $\partial _{i}(\partial _{k}p)=\partial
_{k}(\partial _{i}p)$ and $\partial _{i}(\partial _{k}\upsilon )=\partial
_{k}(\partial _{i}\upsilon )$.

\section{Conclusion and outlook}

In this paper a new look on a classification of the integrable hydrodynamic
chains is presented. The main object which should be under an investigation
is the generating function of conservation laws%
\begin{equation*}
p_{t}=\partial _{x}\psi (\mathbf{u};p),
\end{equation*}%
where all distinct cases are separated by the number $M$ of independent
functions $u^{m}$. The simplest case (\textbf{\ref{b}}) is considered in
details. Each such case has infinitely many sub-cases enumerated by the
number $K$ of independent functions $\upsilon ^{k}$ in the commuting
generating functions of conservation laws%
\begin{equation*}
p_{y}=\partial _{x}\varphi (\mathbf{\upsilon };p).
\end{equation*}%
Thus, all integrable hydrodynamic chains can be split into sub-classes by
virtue of the two numbers $M$ and $K$ only.

Suppose all these functions $\psi (\mathbf{u};p)$ are found. Then the
Gibbons--Tsarev system describing $N$ component hydrodynamic reductions for
every function $\psi (\mathbf{u};p)$ can be derived automatically in the
Riemann invariants and in the field variables $a^{k}$ (which are
conservation law densities) simultaneously. Then corresponding hydrodynamic
type systems with a local Hamiltonian structure (as well as with a priori
prescribed any nonlocal Hamiltonian structure) can be extracted. The
equation of the Riemann surface $\lambda (\mathbf{u};p)$ can be found in
quadratures for the hydrodynamic reductions, whose characteristic velocities
are invariant with respect to any Lie group symmetry. Asymptotic of the
equation of the Riemann surface at the vicinity of any singular point
(usually $\lambda \rightarrow \infty ,p\rightarrow \infty $) yields explicit
expressions of corresponding\ integrable hydrodynamic chains. The ``flat''
hydrodynamic reductions can be used for a construction of a large class of
particular solutions for these hydrodynamic chains and related 2+1
quasilinear equations.

\section*{Acknowledgement}

I thank Eugeni Ferapontov, John Gibbons, Yuji Kodama, Boris Kupershmidt and
Sergey Tsarev for their stimulating and clarifying discussions.

I am grateful to the Institute of Mathematics in Taipei (Taiwan) where some
part of this work has been done, and especially to Derchyi Wu, Jen-Hsu
Chang, Ming-Hien Tu and Jyh-Hao Lee for fruitful discussions.

\end{document}